\documentclass{aa}  
\usepackage{color}
\usepackage{graphicx}
\usepackage{natbib}
\usepackage{txfonts}
\usepackage{lipsum}
\usepackage{subcaption}         
\usepackage{caption} 
\usepackage{lscape}             
\usepackage{placeins}           
\usepackage{color}

\begin{document}

   \title{Narrowing the solar surface flux transport parameter space through nonlinear feedbacks}

   \author{M. Alhosani\inst{1}
        \and M. H. Talafha\inst{2,3}
        \and M. A. Al-Wardat\inst{1,4}
        }

   \institute{Dept. of Applied Physics and Astronomy, University of Sharjah, 27272, Sharjah, UAE\\
             \email{u24103889@sharjah.ac.ae}
            \and Research Institute of Science and Engineering, University of Sharjah, 27272, Sharjah, UAE\\ 
            \and Dept. of Space Physics and Space Technology, HUN-REN Wigner Research Centre for Physics, 1121, Budapest, Hungary.
            \and Sharjah Space and Astronomy Hub, University of Sharjah, 27272, Sharjah, UAE\\ }

   \date{Received MONTH DAY, 2025}

  \abstract
{The surface flux transport (SFT) model describes the evolution of the Sun’s large-scale photospheric magnetic field. While linear transport processes are relatively well constrained, nonlinear feedbacks such as tilt and latitude quenching remain less explored, despite their potential role in regulating the solar cycle.}
{We aim to determine how nonlinear quenching mechanisms and flux decay influence the admissible parameter space of the SFT model, and to identify parameter combinations that reproduce the observed characteristics of the solar polar magnetic field.}
{We extend the parameter-space optimisation by introducing analytic prescriptions for tilt quenching (TQ) and latitude quenching (LQ) in the source term, together with a tunable flux-decay term. The model is solved numerically over a grid of meridional flow speeds ($u_{0}$), surface diffusivities ($\eta$), and decay timescales ($\tau$). Admissible solutions are defined by agreement with {observationally motivated constraints on the {polar-field minimum-to-extremum amplitude ratio}, reversal timing, and the latitude of the polar cap boundary}.}
{{Both TQ and LQ reduce the admissible parameter domain, with LQ exerting the stronger influence. Their combined action produces a pronounced saturation ("ceiling") that limits axial dipole amplification. The inclusion of a finite flux-decay timescale ($\tau \simeq 8$--$10$~yr)
further narrows the admissible domains within the present parameterised source formulation, whereas the non-decaying case may lead to unrealistically persistent dipole fields in this modelling framework. Correlations between $u_{0}$, $\eta$, and $\tau$ reveal a coupled balance between advection, diffusion, and phenomenological flux loss.}}
{{Nonlinear quenching, together with a phenomenological flux-loss timescale, defines a self-limiting SFT regime within the present parameterised framework that constrains dynamo saturation and inherently limits solar-cycle predictability by linking surface transport parameters to nonlinear feedback in the source term.}}


\maketitle
\nolinenumbers
\section{Introduction}
The Surface Flux Transport (SFT) model has successfully reproduced the magnetic flux patterns associated with the Sun’s 22-year magnetic cycle \citep{yeates2023surface}. Incorporating meridional flow, differential rotation, and turbulent diffusion, the model, first formulated by \cite{leighton1964transport}, provides an essential framework for interpreting polar field formation and for predicting the amplitude of future solar cycles through empirical correlations {\citep{munoz2013solar, jiang2018predictability}.}

Key SFT parameters include the meridional flow speed, surface diffusivity, differential rotation, and the decay time associated with radial flux loss. The emergence of bipolar magnetic regions (BMRs), represented through a source term following {\cite{ cameron2007solar}} and implemented as in \cite{petrovay2019optimization}, governs the buildup and reversal of the axial dipole moment \citep{yeates2023surface}. Because these parameters interact nonlinearly and are not independently constrained, their optimisation requires a systematic exploration of parameter space \citep{petrovay2019optimization, whitbread2017parameter}. Uncertainties in BMR emergence properties further complicate this process, as the relationship between sunspot groups and their magnetic flux is not always well constrained \citep{yeo2021relationship}.

Several limitations of classical SFT formulations motivate the need for improved optimisation. These include the reliance on uniform diffusivity and steady, axisymmetric flow profiles, the assumption of linear parameter dependencies, and the neglect of radial diffusion in some models \citep{yeates2023surface}. Such simplifications may introduce errors in predictions of polar field strength and dipole evolution \citep{petrovay2020towards}. Moreover, model sensitivity to meridional flow, tilt properties, and decay time can significantly affect the timing of polar field reversal and the ultimate dipole amplitude \citep{jiang2013can}. {In parameterised SFT formulations based on statistically averaged source terms, the absence of a decay term may lead to unrealistically persistent dipole fields and delayed reversals \citep{petrovay2019optimization}. However, recent data-assimilative simulations that incorporate observed active-region emergence directly have demonstrated that the observed polar-field evolution can be reproduced without invoking an explicit radial diffusion term \citep{yeates2025latitude, wang2025solar}, highlighting the model-dependent nature of this parameterisation.}

Nonlinear feedbacks related to BMR emergence provide an important mechanism for regulating solar cycle amplitude. Observations indicate that both the mean tilt angle and the average emergence latitude vary with cycle strength, resulting in tilt quenching (TQ) and latitude quenching (LQ). These effects reduce the efficiency of dipole buildup in strong cycles and have been examined using SFT and dynamo models \citep{jiang2020nonlinear, talafha2022role}. In coupled dynamo frameworks, such as the 2$\times$2D model of \cite{lemerle2015coupled}, LQ has been shown to play a significant role comparable to TQ, with historical observations (1923–1985) providing stronger evidence for LQ \citep{yeates2025latitude}. The relative importance of these effects also depends on the ratio of meridional flow speed to diffusivity \citep{talafha2022role}.

Previous studies have investigated the nonlinear regulation of the solar cycle within surface flux transport and dynamo frameworks using different approaches. \cite{cameron2010surface} demonstrated that cycle-dependent variations of Joy’s law tilt can effectively limit dipole growth without invoking an explicit decay term. In contrast, \cite{talafha2022role} introduced observable nonlinearities in the form of tilt and latitude quenching and examined their impact on dipole moment modulation. Optimisation studies based on genetic algorithms, such as \cite{lemerle2017coupled}, have identified preferred transport regimes that reproduce key features of the solar cycle, but did not explicitly address how observable nonlinear feedbacks reshape the admissible parameter space. More recently, algebraic treatments have shown that the combined action of tilt and latitude quenching naturally produces a saturation (``ceiling'') of dipole growth with increasing cycle amplitude \citep{talafha2025effect}.

Building on the optimisation framework of \cite{petrovay2019optimization}, the present work incorporates both tilt quenching (TQ) and latitude quenching (LQ) into the surface flux transport (SFT) source term, within a parameter-space optimisation that also includes a finite flux-decay timescale. We quantify how these nonlinear feedbacks reshape the admissible domains of meridional flow speed, surface diffusivity, and decay time, and we assess how their combined action constrains polar field formation {within the present parameterised modelling framework.} Section \ref{sec:method} describes the model setup and the implementation of the quenching prescriptions. The optimisation results are presented in Section \ref{sec:results}. {The interpretation of the results is discussed in Section~\ref{sect:disc} }, and the conclusions are summarised in Section \ref{sec:concl}.

\section{Methodology}
\label{sec:method}

A systematic exploration of the SFT model’s parameter space was performed by \cite{petrovay2019optimization} to optimise its ability to reproduce the solar polar magnetic field. The parameters vary across a wide range of values, and the resulting magnetic field evolution is evaluated against a set of observational constraints, including the timing of polar field reversals, the relative amplitude of the polar field, and the latitudinal structure of the polar field distribution. This approach enables the identification of parameter combinations that produce physically realistic and observationally consistent outcomes.

\subsection{SFT model}
\label{subs:sft}

The evolution of the large-scale radial magnetic field on the solar surface is described by the SFT model through an advective-diffusive transport equation. In its general form, the governing equation is given by
\begin{eqnarray}
 \frac{\partial B}{\partial t} &=& 
 \frac{1}{R\cos{\lambda}}\frac{\partial}{\partial \lambda}(B\,u\,\cos{\lambda})
 \nonumber \\
 &&+\frac{\eta}{R^2\cos{\lambda}}
 \frac{\partial}{\partial \lambda}\left(\cos{\lambda}\frac{\partial B}{\partial \lambda}\right)
 -\frac{B}{\tau} + S(\lambda,t)
\label{eq:sft}
\end{eqnarray}

where $B(\lambda, t)$ is the radial magnetic field as a function of heliographic latitude $\lambda$ and time $t$, $u(\lambda)$ is the meridional flow profile, $\eta$ is the surface diffusivity, $\tau$ is the decay timescale, and $S(\lambda, t)$ represents the source term accounting for flux emergence. The model assumes axial symmetry and a radial field approximation, reducing the problem to one dimension in latitude. This formulation captures the essential transport mechanisms: advection by meridional flow characterised by $u(\lambda)$, carries flux poleward, while differential rotation introduces shear in the longitudinal direction. In the axisymmetric SFT model, longitudinal variations are either averaged out or treated as a background process.  Diffusion represented by $\eta$, accounts for the dispersal of magnetic flux due to supergranular motions. This process helps to smooth out small-scale magnetic features, leading to a more uniform large-scale field distribution. The diffusion coefficient is assumed to be uniform over the solar surface in this work. A decay term $-\frac{B}{\tau}$ is included as a phenomenological representation of vertical diffusion or other three-dimensional effects not explicitly resolved in the model. This term prevents the indefinite accumulation of the global dipole moment by mimicking the loss processes of magnetic flux. The source term, $S(\lambda, t)$, represents the emergence of new magnetic flux, typically through bipolar active regions. It is designed to capture the statistical behaviour of flux emergence over an average solar cycle.

\subsection{Parameter optimization framework} \label{sec:optimization}

To determine the optimal combination of SFT model parameters, we adopt the approach detailed in \cite{petrovay2019optimization}, which involves a systematic, large-scale exploration of the three-dimensional parameter space defined by the meridional flow amplitude ($u_0$), the magnetic diffusivity ($\eta$), and the decay time scale ($\tau$). Rather than relying on algorithmic optimisation techniques such as genetic algorithms, this method evaluates the model's performance across a dense grid of parameter values. This approach involves computing many SFT model realisations across a structured grid of parameter combinations. Each model evolves over a hundred solar cycles until a steady cyclic pattern is established. {The resulting magnetic field configurations are then evaluated against key observationally motivated constraints, including the timing of polar-field reversals relative to sunspot minimum ($T_{\mathrm{rev}}$), {the polar-field minimum-to-extremum amplitude ratio ($R_{\mathrm{pol}}$), defined as $R_{\mathrm{pol}} = {B_{\mathrm{pol}}(t_{\min})}/{B_{\mathrm{pol,ext}}}$, where $B_{\mathrm{pol}}(t_{\min})$ is the polar-field amplitude
at cycle minimum and $B_{\mathrm{pol,ext}}$ is the extremal (peak absolute) polar-field value during the cycle}, the latitude of the polar cap boundary ($\lambda_{\mathrm{cap}}$). In addition, we consider constraints on the axial dipole moment evolution, namely the dipole reversal timing ($D_{\mathrm{rev}}$) and {the axial dipole moment minimum-to-extremum amplitude ratio ($D_{\mathrm{min/ext}}$), defined as
$D_{\mathrm{min/ext}} ={D(t_{\min})}/{D_{\mathrm{ext}}}$}. The first three constraints are used to define the primary admissible parameter domains, while the dipole-based constraints provide complementary diagnostics of the global magnetic-field evolution. This separation allows us to distinguish between constraints directly linked to observable polar-field properties and those probing the global dipole evolution, thereby providing a more comprehensive assessment of the SFT parameter space.} Models that satisfy all observational criteria are classified as admissible, and the corresponding parameter sets are identified as optimal within the context of the selected meridional flow profile. This approach allows for a transparent delineation of allowed and excluded regions in parameter space and offers clear physical insight into the role of each parameter in shaping solar magnetic field evolution.

To ensure that the optimisation results are not biased by the peculiarities of any specific cycle, the source term $S(\lambda, t)$ is constructed to represent a statistically averaged solar cycle. It is modulated as a smooth, axisymmetric distribution that captures the net effect of tilted bipolar magnetic region (BMR) emergence. It consists of pairs of Gaussian flux rings of opposite polarity, whose latitudinal positions and separations evolve throughout the cycle according to empirical fits derived from observational studies. The amplitude and temporal profile of the source are based on the cycle-averaged sunspot emergence rate, ensuring that the resulting field evolution reflects typical solar behaviour rather than cycle-specific anomalies (the exact formulation is given in \cite{petrovay2019optimization}). 

To incorporate nonlinear effects into the source term of the SFT model, we follow the formulation described by \cite{talafha2022role}, where the emergence properties of active regions are modulated by the solar cycle amplitude. Two observable nonlinearities are introduced: TQ and LQ. In the case of TQ, the latitudinal separation $\Delta\lambda$ of the bipolar magnetic regions (representing Joy’s law) is reduced for stronger cycles, following the relation

\begin{equation}
\Delta\lambda = 1^\circ.5 \sin\lambda_0 \left(1 - b_{\mathrm{joy}} \frac{A_n - A_0}{A_0}\right),
\end{equation}

where $A_n$ is the amplitude of the $n$th cycle, $A_0$ is the reference amplitude for an average cycle, and $b_{\mathrm{joy}}$ is the nonlinearity parameter controlling the strength of TQ with a reference value of 0.15.

For LQ, the mean latitude $\lambda_0$ of flux emergence is increased with cycle amplitude according to the empirical formula
\begin{equation}
\lambda_n = 14^\circ.6 + b_{\mathrm{lat}} \frac{A_n - A_0}{A_0},
\end{equation}
where $b_{\mathrm{lat}} \approx 2.4$ is derived from observations \citep{jiang2011solar}. This modified mean latitude feeds into a time-dependent latitudinal profile,
\begin{equation}
\lambda_0(t; n) = \left[26.4 - 34.2\left(\frac{t}{P}\right) + 16.1\left(\frac{t}{P}\right)^2\right] \left(\frac{\lambda_n}{14^\circ.6}\right),
\end{equation}
where $P$ is the cycle period. Both effects alter the spatiotemporal structure of the source term $S(\lambda, t)$, leading to cycle-dependent variations in flux emergence characteristics.

To evaluate the realism of each SFT model realisation, we compare its output against a set of well-established observational constraints that characterise the solar polar magnetic field {and dipole moment} during a typical cycle. Specifically, $T_{rev}$ must fall within observed ranges, typically around 3.4–4.3 years after the minimum, and the field amplitude ratios must reflect the gradual decline of polar field strength after its peak. Additionally, $\lambda_{\mathrm{cap}}$ is constrained by the requirement that the field drops to half its polar value between latitudes $65^\circ$ and $75^\circ$. These criteria, summarised in Table~1 of \cite{petrovay2019optimization}, provide quantitative benchmarks for model validation. Parameter combinations that produce magnetic field evolutions satisfying all of these constraints are classified as admissible. 

The numerical implementation of the SFT model follows a one-dimensional explicit time-stepping scheme, designed to ensure magnetic flux conservation throughout the simulation. The model is run on a latitudinal grid with a resolution of $0.5^\circ$, which is sufficient to accurately capture the structure of the polar cap field and meet the constraints on the topknot width. A fixed timestep of 6 hours is used, ensuring numerical stability across all parameter combinations tested. Each simulation is initialised with a dipolar field configuration and evolved over 100 solar cycles, allowing transient effects to decay and a quasi-steady cyclic behaviour to emerge. Once the model reaches steady oscillations, the results from the final cycle are evaluated against the observational constraints.

\section{Results}
\label{sec:results}

\begin{figure}[t]
	\centering
	\includegraphics[width=1.0\linewidth]{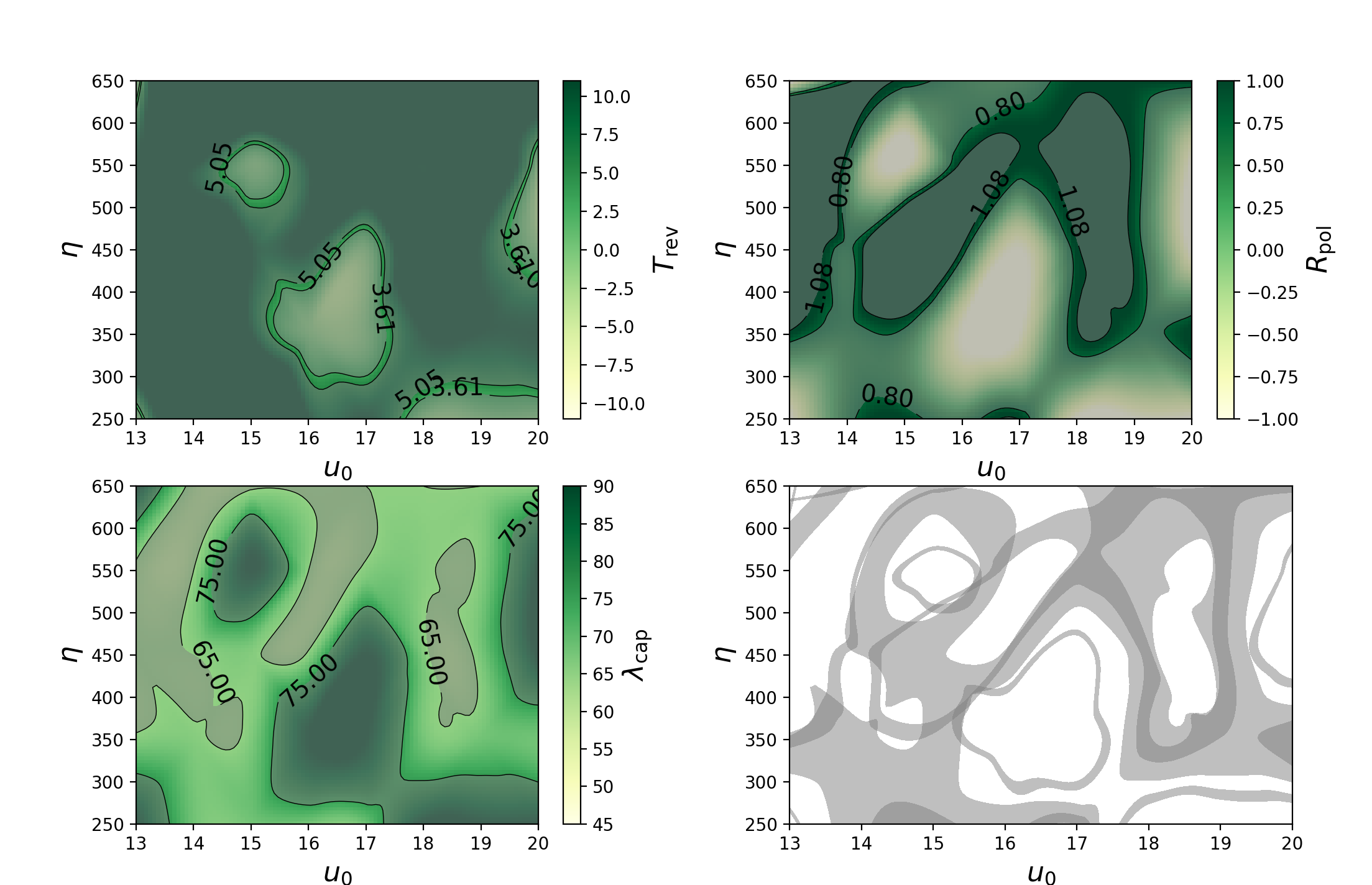}
	\caption{Admissible domains in the $(u_{0}, \eta)$ parameter space for the case without tilt or latitude quenching, with a decay timescale of $\tau = 8$~yr. The coloured maps indicate the parameter domains that {satisfy the three model constraints: the polar-field reversal timing ($T_{\mathrm{rev}}$), {the polar-field minimum-to-extremum amplitude ratio} ($R_{\mathrm{pol}}$), and the polar cap boundary latitude ($\lambda_{\mathrm{cap}}$). The numerical annotations denote the $\pm1\sigma$ intervals associated with each constraint, following the optimisation framework of \citet{petrovay2019optimization}}. The grey-shaded region shows the intersection of these constraints, corresponding to parameter combinations that simultaneously satisfy all three criteria. This case serves as the reference configuration for comparison with the quenching models shown in Figs.~\ref{fig:TQ_0.15_8}--\ref{fig:LQTQ_1000}.}
    \label{fig:noq}
\end{figure}

To investigate the impact of latitude and tilt quenching on the admissible parameter space, we first examine the case with no quenching applied (linear case), focusing on how the accepted regions are constrained by the three constraints mentioned earlier. For these tests, the decay timescale was fixed at $\tau = 8$~yr. Figure~\ref{fig:noq} shows the results for the linear case: the admissible domain is concentrated at $\eta$ values between 600 and 650 $\mathrm{km^{2}\,s^{-1}}$ when $u_{0}$ lies between 18 and 19 $\mathrm{m\,s^{-1}}$, with an additional smaller region appearing around $\eta \approx 250$ and $u_{0}$ between 14 and 15 $\mathrm{m\,s^{-1}}$. These baseline regions provide a useful reference for identifying how the parameter space is modified once quenching mechanisms are introduced in subsequent simulations.

\begin{figure}[t]
	\centering
	\includegraphics[width=1.0\linewidth]{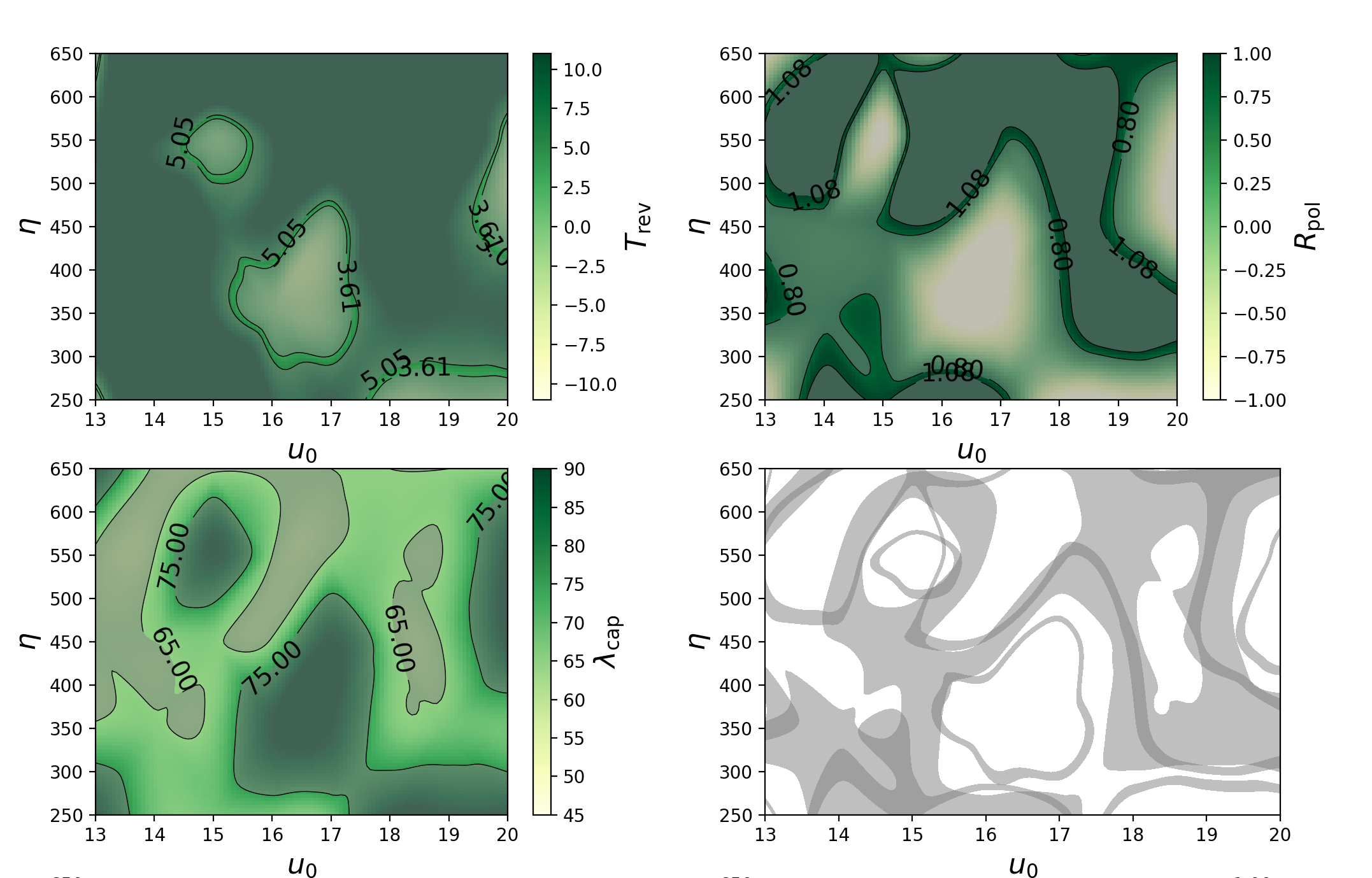}
	   \caption{Same as Figure~\ref{fig:noq} for the case with TQ included ($b_{\mathrm{joy}} = 0.15$) only, with a decay timescale of $\tau = 8$~yr. Compared to the no-quenching case, the effect of TQ is modest, leading to only slight contraction of the admissible domains.}
    \label{fig:TQ_0.15_8}
\end{figure}

\begin{figure}[t]
	\centering
	\includegraphics[width=1.0\linewidth]{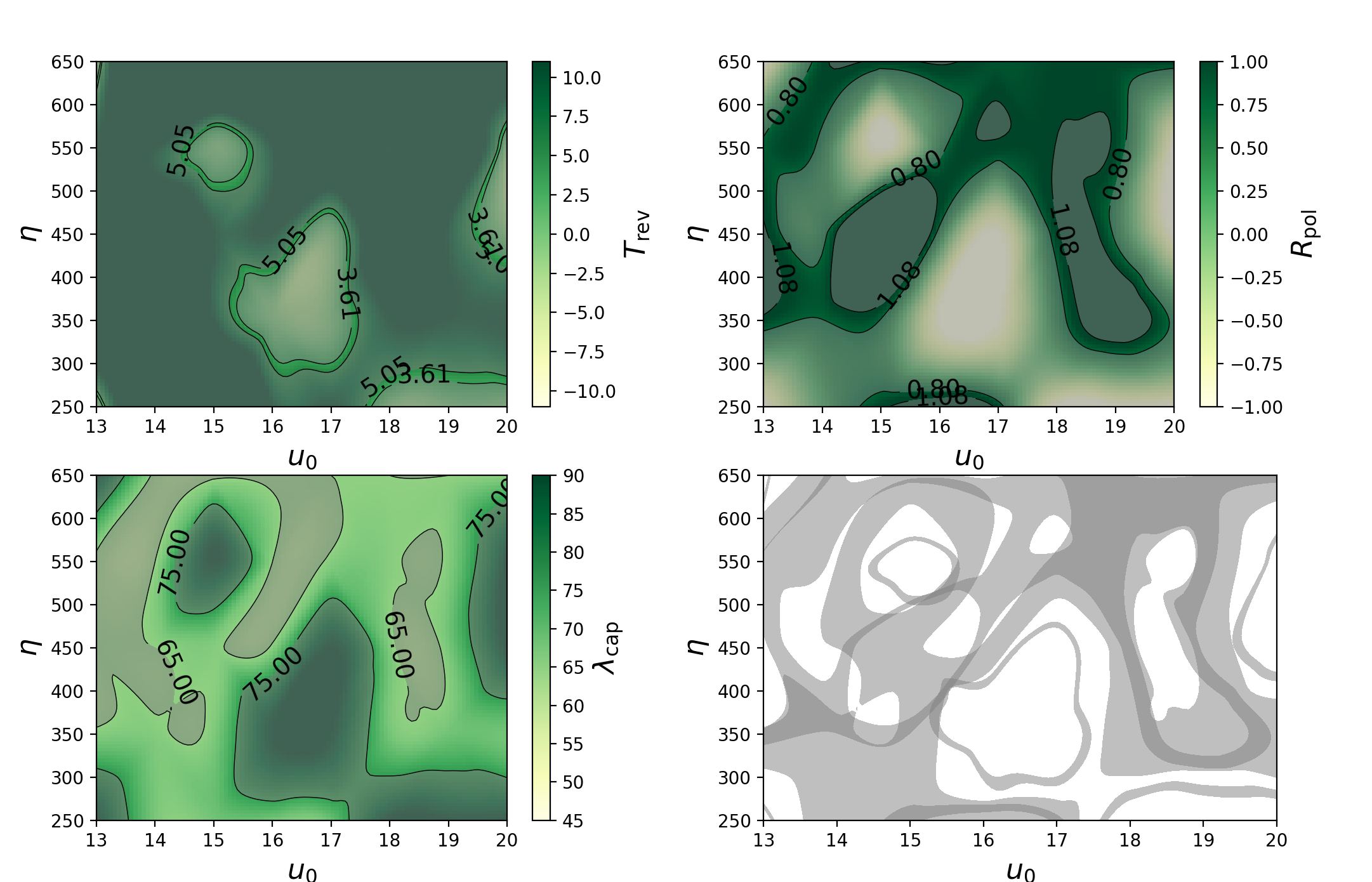}
	\caption{Same as Figure~\ref{fig:noq} for the case with LQ included ($b_{\mathrm{lat}} = 2.4$) only, with a decay timescale of $\tau = 8$~yr. In contrast to the TQ case (Fig.~\ref{fig:TQ_0.15_8}), LQ has a stronger impact, narrowing the admissible parameter ranges for both meridional flow speed and diffusivity.}
    \label{fig:LQ_2.4_8}
\end{figure}

\begin{figure}[t]
    \centering
    \includegraphics[width=1.0\linewidth]{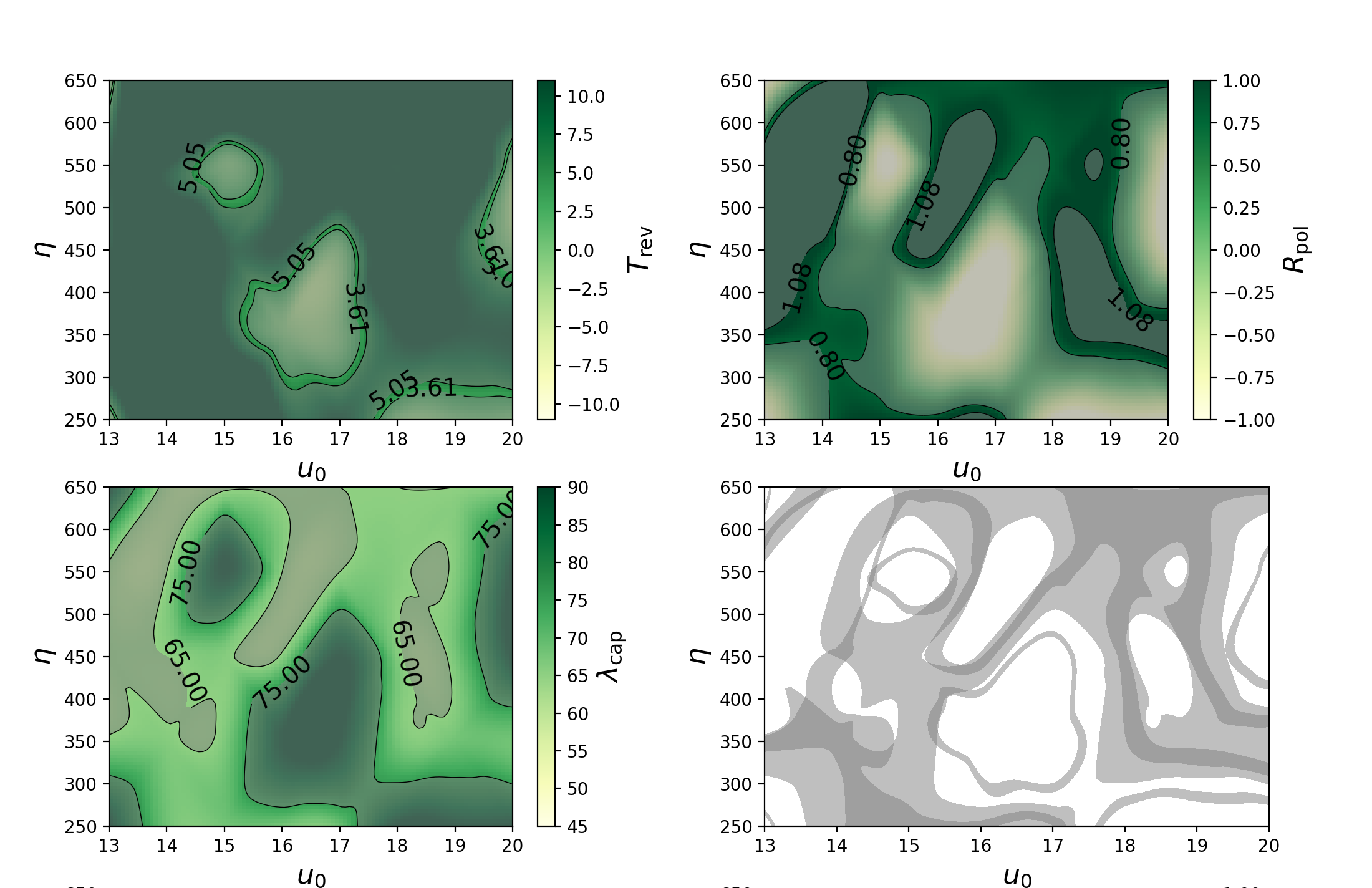}
     \caption{Same as Figure~\ref{fig:noq} for the case with both TQ ($b_{\mathrm{joy}} = 0.15$) and LQ ($b_{\mathrm{lat}} = 2.4$) included, with a decay timescale of $\tau = 8$~yr. Compared to the LQ and TQ cases (Figs.~\ref{fig:TQ_0.15_8} and \ref{fig:LQ_2.4_8}), the combined effect of tilt and latitude quenching significantly restricts the admissible domains, shifting them toward narrower bands in $u_{0}$ and $\eta$.}
    \label{fig:LQTQ_8}
\end{figure}

Figure~\ref{fig:TQ_0.15_8} illustrates the case {where} only TQ is applied, which highlights the isolated effect of TQ on the admissible parameter space. In contrast, Figure~\ref{fig:LQ_2.4_8} shows the influence of LQ alone, where the best-fit parameter values are concentrated around $u_{0} = 17$--$19$ $\mathrm{m\,s^{-1}}$ and $\eta \approx 650$ $\mathrm{km^{2}\,s^{-1}}$. When both nonlinearities are included simultaneously, Figure~\ref{fig:LQTQ_8}, the admissible domain becomes even more restricted, with allowed values clustering near $u_{0} = 18$--$19$ $\mathrm{m\,s^{-1}}$ and $\eta = 600$--$650$ $\mathrm{km^{2}\,s^{-1}}$. 

\begin{figure}[t]
	\centering
	\includegraphics[width=1.0\linewidth]{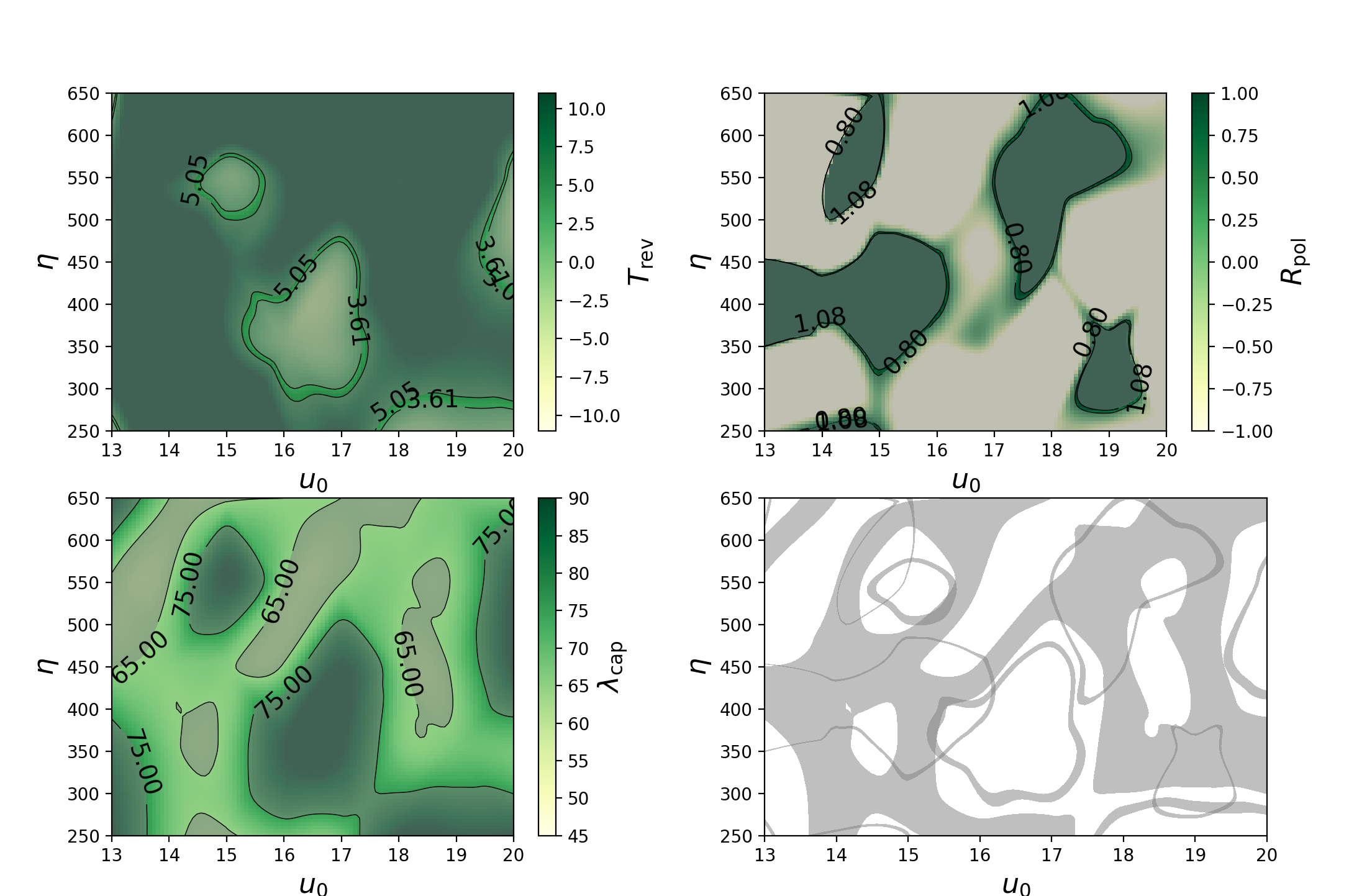}
	\caption{Same as Figure~\ref{fig:noq} for the case with both TQ ($b_{\mathrm{joy}} = 0.15$) and LQ ($b_{\mathrm{lat}} = 2.4$) included, with a long decay timescale of $\tau = \infty$. Compared to the short decay case (Fig.~\ref{fig:LQTQ_8}), the admissible domain expands considerably, illustrating that the absence of effective flux decay relaxes the parameter constraints, although such models tend to produce unrealistically delayed dipole reversals.}
    \label{fig:LQTQ_1000}
\end{figure}

For a longer decay timescale of $\tau = \infty$, Figure~\ref{fig:LQTQ_1000} presents the case with both quenching mechanisms. Compared to the short-decay case in Figure~\ref{fig:LQTQ_8} for $\tau = 8$~yr, the admissible grey-shaded region is considerably larger, demonstrating how the absence of effective decay broadens the parameter space. 

In the absence of quenching, the optimal values for $\eta$ lie between 300--350 $\mathrm{km^{2}\,s^{-1}}$ and around 650 $\mathrm{km^{2}\,s^{-1}}$, with $u_{0}$ concentrated between 16 and 17 $\mathrm{m\,s^{-1}}$. When only TQ is included, the best values remain nearly identical to the linear case, which shows that TQ {has only} a modest effect on the parameter space. By contrast, LQ produces a broader admissible range, with $\eta$ spanning 300--350 and 600--650 $\mathrm{km^{2}\,s^{-1}}$, and $u_{0}$ extending from 16 to 18 $\mathrm{m\,s^{-1}}$. When both quenching mechanisms are applied simultaneously, the admissible domain narrows again, with $\eta$ clustering near 350 and 650 $\mathrm{km^{2}\,s^{-1}}$, while $u_{0}$ is restricted to the lower range of 14--16 $\mathrm{m\,s^{-1}}$.

\begin{table*}[hbtp]
\caption{\label{tbl:const}{Summary of admissible ranges of the meridional flow speed $u_{0}$ (in $\mathrm{m\,s^{-1}}$) and surface diffusivity $\eta$ (in $\mathrm{km^{2}\,s^{-1}}$) for different quenching configurations.}}
\centering
\begin{tabular}{llllll}
\hline\hline
$\tau$ [yr] &[$b_{joy}$,$b_{lat}$] &{$T_{\mathrm{rev}}$} &{$R_{\mathrm{pol}}$}  &{$\lambda_{\mathrm{cap}}$}  &{Admissible ranges}\\
\hline
8&[0, 0] & $\eta$= 300-450  & $\eta$= 650 & $\eta$= 300-350; 600-650 & $\eta$= 300-350; 650\\
& & $u_{0}$ = 16-17  & $u_{0}$ = 18-19  &  $u_{0}$ = 14-15; 18-19  & $u_{0}$ = 16-17 \\
\hline
8&[0.15, 0]  & $\eta$= 300-450  & $\eta$= 650; 250-350  & $\eta$= 300-350; 600-650 & $\eta$= 300-350; 650\\
& & $u_{0}$ = 16-17  & $u_{0}$ = 14; 15  &  $u_{0}$ = 14-15; 18-19  & $u_{0}$ = 16-17 \\
\hline
8& [0, 2.4]& $\eta$= 300-450  & $\eta$= 650; 500-650  & $\eta$= 300-350; 600-650  & $\eta$= 300-350; 600-650\\
& & $u_{0}$ = 16-17 & $u_{0}$ = 15-20; 16-18 &  $u_{0}$ = 14-15; 18-19 & $u_{0}$ = 16-18\\
\hline
8& [0.15, 2.4]& $\eta$= 300-450 & $\eta$= 350; 650; 300 & $\eta$= 300-350; 600-650 & $\eta$=350; 650\\
& & $u_{0}$ = 16-17 & $u_{0}$ = 14; 14-19; 18-19 &  $u_{0}$ = 14-15; 18-19 & $u_{0}$ = 14-16\\
\hline
\hline

$\infty$& [0, 0] & $\eta$= 300-450 & $\eta$= 400-450; 650 & $\eta$= 250-650 & $\eta$= 300-450; 550-650\\
& & $u_{0}$ = 16-17 & $u_{0}$ = 16-17; 18 &  $u_{0}$ = 14-16; 18-20  & $u_{0}$ = 16-17\\
\hline
$\infty$& [0.15, 0]  & $\eta$= 300-450 & $\eta$= 550-650 & $\eta$= 250-650 & $\eta$= 300-450; 550-650\\
& & $u_{0}$ = 16-17 & $u_{0}$ = 19-20 &  $u_{0}$ = 14-16; 18-19 & $u_{0}$ = 14-20\\
\hline
$\infty$& [0, 2.4]& $\eta$= 300-450 & $\eta$= 550-650 & $\eta$= 250-450; 500-650 & $\eta$= 300-450; 550-650\\
& & $u_{0}$ = 16-17 & $u_{0}$ = 14 &  $u_{0}$ = 14-15; 18-19 & $u_{0}$ = 14-19\\
\hline
$\infty$ & [0.15, 2.4]& $\eta$= 300-450 & $\eta$= 550-600 & $\eta$= 300-450; 600-650 & $\eta$=300-450\\
& & $u_{0}$ = 16-17 & $u_{0}$ = 19 &  $u_{0}$ = 14-15; 18-19 & $u_{0}$ = 14-19\\
\hline
\end{tabular}
\tablefoot{Each row corresponds to a distinct case: no quenching, tilt quenching (TQ; $b_{\mathrm{joy}}=0.15$), latitude quenching (LQ; $b_{\mathrm{lat}}=2.4$), and the combined TQ+LQ model. The listed intervals represent the ranges of parameters that simultaneously satisfy the three model constraints: the polar-field reversal timing ($T_{\mathrm{rev}}$), {the polar-field minimum-to-extremum amplitude ratio} ($R_{\mathrm{pol}}$), and the polar cap boundary latitude ($\lambda_{\mathrm{cap}}$). The table illustrates how the admissible parameter domains shift and contract depending on the inclusion of nonlinear quenching mechanisms.}
\end{table*}

The inclusion of quenching mechanisms has a clear physical impact on the admissible domains of the $u_{0}-\eta$ parameter space. When no quenching is applied, the allowed regions are relatively broad, with multiple disjoint islands of admissibility. Introducing TQ alone produces only a modest contraction of these regions, indicating that the suppression of Joy’s law tilt does not drastically alter the large-scale transport balance. In contrast, LQ exerts a stronger influence. By systematically shifting flux emergence toward higher latitudes in stronger cycles, it enhances the poleward cancellation of flux and thereby narrows the admissible windows of both $u_{0}$ and $\eta$. When both nonlinearities act together, the admissible domains shrink further and shift toward lower meridional flow amplitudes, resulting in a more selective parameter space that reflects the "ceiling effect" reported in recent algebraic approaches \citep{talafha2025modelling}. This ceiling arises because additional flux input no longer translates into stronger polar fields once quenching limits the efficiency of dipole build-up.

{The $T_{\mathrm{rev}}$} constraint shows only a weak dependence on the inclusion of TQ or LQ and remains primarily governed by the balance between meridional flow speed and the effective flux-decay timescale. By contrast, the {$R_{\mathrm{pol}}$} is strongly affected by nonlinear feedbacks, with latitude quenching producing the most pronounced reduction of the admissible parameter ranges. The {$\lambda_{\mathrm{cap}}$} exhibits comparatively little sensitivity to source-term modulation across all model configurations, reflecting its predominantly geometrical character. When all three constraints are applied simultaneously, the intersection of admissible ranges contracts substantially under combined TQ and LQ, yielding a markedly narrower region of parameter space than in the unquenched case.

The role of the decay term is equally important. With a short decay timescale ($\tau=8$~yr), corresponding to strong radial diffusion, the model requires tightly constrained values of $\eta$ and $u_{0}$ to satisfy the observational benchmarks, and the combined effect of quenching mechanisms sharply limits the admissible domains. In contrast, when the decay term is effectively absent ($\tau=\infty$), the admissible parameter space expands considerably, even under both tilt and latitude quenching. Physically, this reflects that without significant flux loss, the system can sustain a broader range of transport conditions while still meeting the polar field constraints. 
{In parameterised SFT models based on statistically averaged source terms, the absence of a decay term may lead to unrealistically persistent dipole fields and delayed reversals. However, recent data-assimilative SFT simulations that incorporate observed active-region emergence directly (e.g. Yeates et al.~2025; Wang et al.~2025) have demonstrated that the observed polar-field evolution can be reproduced without invoking an explicit radial diffusion term.}

{It is therefore important to emphasise that the flux-decay timescale $\tau$ introduced in the present optimisation framework should not be interpreted as a literal physical radial diffusivity. Instead, it serves as a phenomenological parameter that represents unresolved three-dimensional processes, such as turbulent radial mixing or flux submergence below the photosphere, which are not explicitly captured by surface-only transport models. In SFT simulations that assimilate observed active regions, such loss processes may be implicitly accounted for through the source term itself, potentially reducing or eliminating the need for an explicit decay term \citep{yeates2025latitude, wang2025solar, luo2025simulation}. The optimisation results presented here should therefore be understood as conditional on the adopted parameterised source formulation.}

{This model dependence is further highlighted by recent developments
in alternative numerical realisations of the SFT equation. In particular, \cite{Athalathil2026PINN} employed a Physics-Informed Neural Network (PINN) framework to solve the SFT equation in a mesh-free setting, incorporating nonlinear tilt and latitude quenching without invoking an explicit radial diffusion or decay term. Despite the absence of such a loss term, their simulations reproduced the observed cycle-to-cycle modulation of the polar field, suggesting that the need for a decay term may depend on the numerical implementation of nonlinear feedback mechanisms. This supports the interpretation that the admissible parameter domains identified in the present optimisation study are conditional on the adopted parameterised source formulation and may not represent a unique physically required regime.}

\begin{figure}[t]
	\centering
	\includegraphics[width=1.0\linewidth]{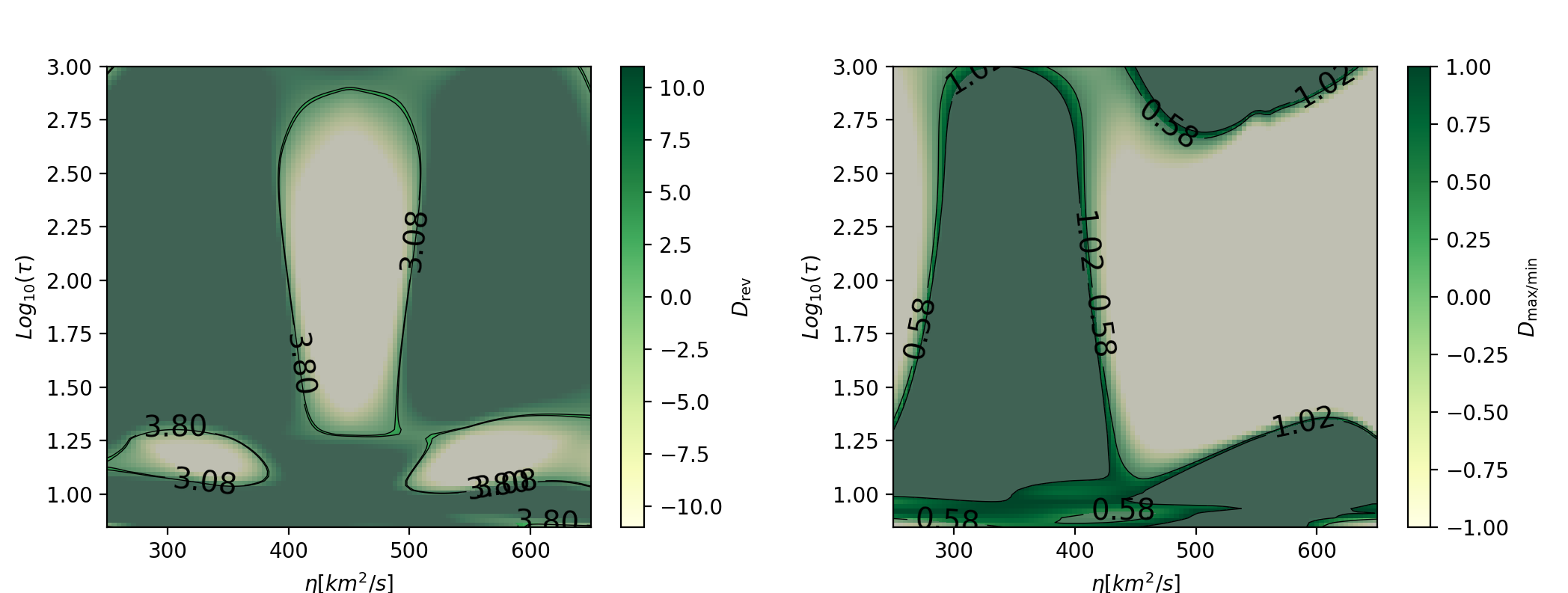}
    \caption{{Dependence of the admissible parameter domain on the decay timescale, expressed as $\log_{10}(\tau)$, plotted against the surface diffusivity $\eta$ for the case including both latitude and tilt quenching ($b_{\mathrm{lat}} = 2.4$, $b_{\mathrm{joy}} = 0.15$). These panels illustrate the dipole-based constraints on the global magnetic-field evolution. Left: axial dipole moment reversal time ($D_{\mathrm{rev}}$) as a function of $\eta$ and $\log_{10}(\tau)$. Right: normalised {axial dipole moment minimum-to-extremum amplitude ratio ($D_{\mathrm{min/ext}}$)}. The colour-coded contours show how the decay timescale modulates both the timing and amplitude of the global dipole moment: shorter decay timescales (lower $\tau$) restrict the admissible domain to lower diffusivities, whereas longer decay timescales broaden the range of permissible $\eta$.}}
    \label{fig:u0_12_LQTQ}
\end{figure}

\begin{figure}[t]
	\centering
	\includegraphics[width=1.0\linewidth]{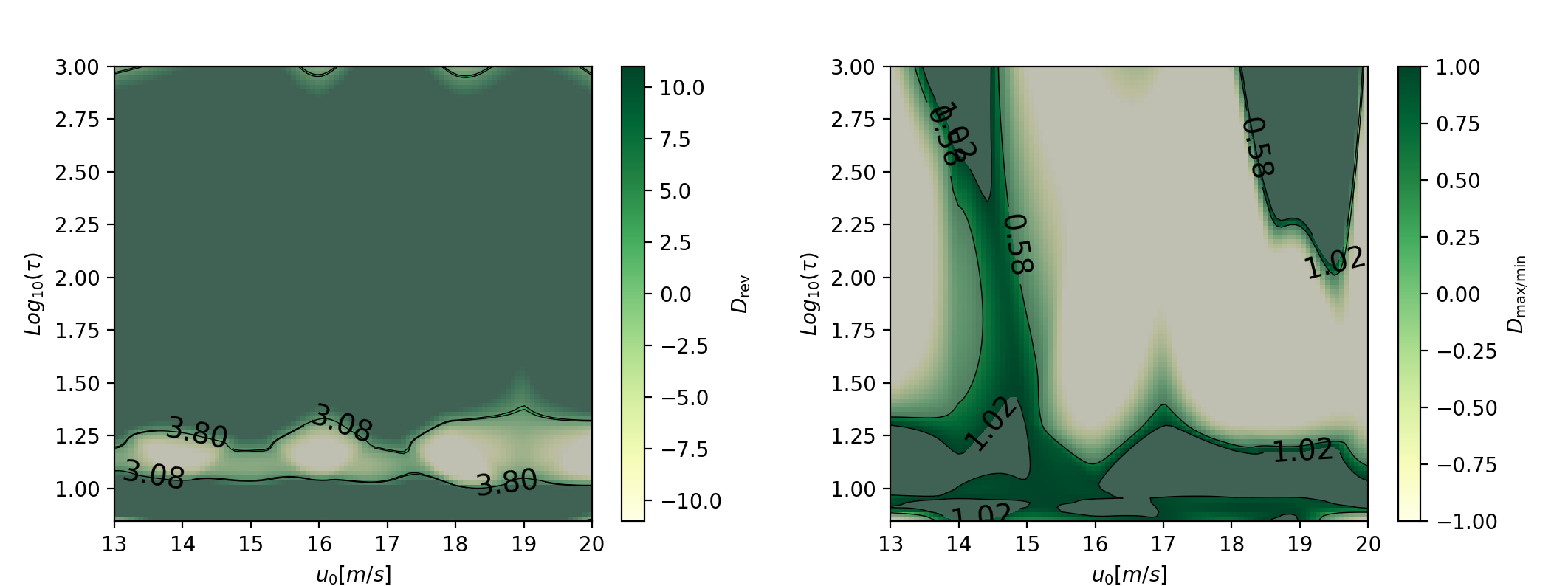}
	\caption{{Dependence of the admissible parameter domain on the decay timescale, expressed as $\log_{10}(\tau)$, plotted against the meridional flow amplitude $u_{0}$ for the case including both latitude and tilt quenching ($b_{\mathrm{lat}} = 2.4$, $b_{\mathrm{joy}} = 0.15$), at a fixed diffusivity of $\eta = 450~\mathrm{km^{2}\,s^{-1}}$. These panels illustrate the dipole-based constraints on the global magnetic-field evolution. Left: axial dipole moment reversal time ($D_{\mathrm{rev}}$) as a function of $u_{0}$ and $\log_{10}(\tau)$. Right: normalised {axial dipole moment minimum-to-extremum amplitude ratio ($D_{\mathrm{min/ext}}$)}. The colour-coded contours show that, for a fixed diffusivity, shorter decay timescales (lower $\tau$) restrict the admissible range to slower meridional flows, whereas longer decay timescales allow higher $u_{0}$.}}
    \label{fig:eta_450_LQTQ}
\end{figure}

Figures~\ref{fig:u0_12_LQTQ} and \ref{fig:eta_450_LQTQ} illustrate how the admissible domains vary with the decay timescale, expressed as $\log_{10}(\tau)$, in the presence of both nonlinear quenching mechanisms. In Fig.~\ref{fig:u0_12_LQTQ}, $\log_{10}(\tau)$ is plotted against diffusivity $\eta$, showing that longer decay timescales (weaker decay) broaden the admissible range, while shorter $\tau$ values confine it to narrower intervals. Figure~\ref{fig:eta_450_LQTQ} presents the complementary case, where $\log_{10}(\tau)$ is plotted against the meridional flow amplitude $u_{0}$. For a fixed diffusivity, the admissible $u_{0}$ values shift toward lower ranges as $\tau$ decreases, indicating that stronger flux decay can be compensated by slower poleward transport. Together, these results confirm that, when quenching mechanisms are active, $\tau$, $\eta$, and $u_{0}$ become tightly coupled, defining a narrow surface of admissible solutions. This contrasts with the broader parameter domains identified by \citet{petrovay2019optimization}, where the absence of nonlinear quenching allowed a wider range of $(u_{0}, \eta)$ combinations for $\tau \simeq 5$--10~yr. The present results, therefore, extend their framework by showing that the inclusion of tilt and latitude quenching further constrains the physically viable region, particularly for shorter decay timescales.

Figure~\ref{fig:eta_450_LQTQ}, which examines the case at $\eta = 450~\mathrm{km^{2}\,s^{-1}}$ as a representative cut, demonstrates that admissibility collapses to a narrow band in $u_{0}$ once nonlinear feedbacks are included. For fixed diffusivity, only a limited range of meridional flow speeds can counterbalance the suppression effects of tilt and latitude quenching while maintaining agreement with observational constraints. This nonlinear coupling is consistent with, but more restrictive than, the results of \citet{petrovay2019optimization}, who found that models with $\tau \gtrsim 10$~yr and no quenching reversed the dipole too late in the cycle. In contrast, acceptable solutions existed for $\tau \sim 5$--10~yr with higher $\eta$ values (500--800~$\mathrm{km^{2}\,s^{-1}}$). In the quenched models presented here, the same $\tau$ range remains optimal, yet the admissible $(u_{0}, \eta)$ combinations are significantly reduced, forming narrow ridges rather than extended ``islands'' in parameter space. This demonstrates that nonlinear quenching mechanisms reinforce the decay constraint identified by \citet{petrovay2019optimization}, yielding a more self-limiting and physically consistent configuration for solar cycle modulation.

In this context, the decay timescale $\tau$ plays a dual role that merits clarification. Physically, it is commonly interpreted as a parametrisation of vertical diffusion or other three-dimensional processes not explicitly captured in surface flux transport models, such as flux submergence or radial turbulent mixing. Within the present optimisation framework, however, $\tau$ should be regarded primarily as a phenomenological control that limits the long-term memory of the surface field and prevents the accumulation of unrealistically persistent dipole moments. The preferred range $\tau \simeq 8$--$10$~yr therefore reflects the effective timescale required for the SFT model to reproduce observed polar-field reversal timing in the presence of nonlinear quenching. Importantly, the inclusion of tilt and latitude quenching does not eliminate the need for a finite decay term; instead, both effects act together to further restrict the admissible parameter space and enforce a physically self-consistent, self-limiting regime of polar field evolution.

\section{Discussion}
\label{sect:disc}
The admissible domains of $\eta$ and $u_{0}$ listed in Table~\ref{tbl:const} can be directly compared with the earlier parameter-space study of \citet{petrovay2019optimization}. In their work, the inclusion of a decay term was found to be essential for reproducing realistic polar field reversals: models without decay produced excessively late reversals, even if the amplitude constraints could be satisfied. Our results {are consistent with} this conclusion by showing that the admissible parameter space is much broader when $\tau \rightarrow \infty$ as seen in Fig.~\ref{fig:LQTQ_1000}, whereas it becomes sharply restricted for $\tau = 8$~yr (Figs.~\ref{fig:noq} \-- \ref{fig:LQTQ_8}).

The present optimisation considers both nonlinearities: TQ alone introduces only minor modifications compared to the linear case, while LQ exerts a stronger influence by narrowing the admissible domains of both $\eta$ and $u_{0}$. When both quenching mechanisms are included, the admissible domains converge toward the narrow bands already identified by \citet{petrovay2019optimization}, suggesting that the combination of finite decay and nonlinear feedbacks naturally drives the system toward restricted parameter regimes. This provides a physical interpretation for the ``islands'' of admissibility previously mapped, which can be understood as the combined outcome of decay-driven flux loss and quenching-driven flux suppression.

The quenching parameters adopted in this study, $b_{\rm joy}=0.15$ for tilt quenching and $b_{\rm lat}=2.4$ for latitude quenching, are fixed at values motivated by observational and modelling studies \citep{jiang2011solar,talafha2022role}. While a systematic exploration of the sensitivity to these parameters is beyond the scope of the present work, we note that moderate variations in their amplitudes primarily affect the quantitative extent of the admissible domains rather than their qualitative structure. In particular, the contraction of the admissible parameter space, the stronger influence of latitude quenching relative to tilt quenching, and the emergence of a saturation (``ceiling'') effect in dipole amplification are robust features that persist for physically reasonable choices of the quenching strengths. The fixed-parameter approach adopted here, therefore, suffices to assess how observable nonlinear feedbacks reshape the SFT optimisation landscape.

A direct comparison can be made with the surface flux transport simulations of \citet{cameron2010surface}, who modelled solar cycles~15--21 by incorporating the observed {active-region properties into the source term. In addition to the cycle-to-cycle variations of Joy’s law tilt angles, their use of observed emergence latitudes implicitly introduced latitude-dependent modulation of the source term}. Their model successfully reproduced the timing of the {polar-field} reversals and the amplitude of the open flux without requiring any additional decay term, demonstrating that the empirically observed anti-correlation between cycle strength and mean tilt angle, {together with cycle-dependent emergence latitudes,} can provide an effective self-regulating mechanism. In contrast, the present optimisation study introduces both tilt and latitude quenching as explicit nonlinear functions in the {parameterised} source term, together with an adjustable flux-decay timescale $\tau$. The results confirm the central conclusion of \citet{cameron2010surface}, that nonlinear suppression of tilt plays a critical role in limiting dipole growth, but further show that, {within this parameterised framework,} the inclusion of latitude-dependent quenching and flux decay substantially tightens the admissible $(u_{0},\eta,\tau)$ domains. Whereas \citet{cameron2010surface} achieved stable reversals over a broad parameter range using {observationally constrained emergence properties}, our results indicate that when both quenching mechanisms and decay are accounted for {in a statistically averaged source formulation}, the acceptable solutions cluster along narrow ridges in parameter space.

A meaningful comparison can also be made with the optimisation results of \citet{lemerle2017coupled}, who employed a genetic algorithm to calibrate a two-dimensional flux-transport dynamo model against observed solar cycle features. Their study identified an optimal regime characterized by a diffusivity of $\eta \simeq 450$--$600~\mathrm{km^{2}\,s^{-1}}$ and a meridional flow amplitude of $u_{0} \simeq 12$--$18~\mathrm{m\,s^{-1}}$, which produced the best agreement with the timing and amplitude of the solar dipole reversals. The parameter ranges obtained in the present study are broadly consistent with those results, particularly for models including moderate decay timescales ($\tau \sim 8$--10~yr). However, when nonlinear tilt and latitude quenching mechanisms are introduced, the admissible domains shrink markedly and align along narrow ridges in the $(u_{0}, \eta)$ plane, rather than the broader basins of attraction reported by \citet{lemerle2017coupled}. This indicates that while both optimisation approaches converge on similar transport amplitudes, the inclusion of nonlinear feedback enforces tighter coupling between advection, diffusion, and decay processes. In this sense, the present model extends the empirical optimisation of \citet{lemerle2017coupled} by providing a physically constrained explanation for the restricted range of solutions that reproduce observed solar cycle variability.

The inclusion of explicit nonlinear quenching mechanisms provides a physically grounded explanation for the self-regulation of the solar cycle amplitude. Tilt and latitude quenching together reproduce the observed saturation of the axial dipole moment, linking flux-transport parameters to the nonlinear backreaction of magnetic activity on the source term. The resulting narrow admissible domains suggest that the solar cycle operates near a marginally stable regime, where modest variations in surface flow or diffusivity can produce significant modulation in dipole strength. This reinforces the interpretation of the solar dynamo as a weakly nonlinear system, consistent with the empirical correlations between tilt angle, emergence latitude, and cycle strength reported in observational studies.

From the perspective of solar cycle predictability, these results imply that the dynamo operates close to a saturation threshold. When quenching feedbacks are active, small perturbations in surface transport or emergence statistics can lead to disproportionately large changes in polar field buildup, thereby limiting the predictive horizon of purely kinematic models. The quenching mechanisms thus provide a natural explanation for the stochastic component of solar cycle variability: while nonlinear feedbacks constrain the mean cycle amplitude, the detailed timing and asymmetry of reversals may remain sensitive to fluctuations in active region emergence. 

{We also note that the adopted cycle profile follows the analytic
formulation of a quadratic fit derived by \cite{jiang2011solar} from many observed
solar cycles. Although this profile provides a useful statistically averaged description of solar cycle emergence, observed solar cycles are not strictly self-similar. In particular, \citet{jiang2018predictability} showed that the rising phase is cycle-dependent, whereas the declining phase is comparatively less variable. The use of a simplified cycle profile may therefore influence the inferred admissible parameter ranges, and exploring the impact of more realistic cycle-dependent source profiles will be an important direction for future work.}

\section{Conclusion}
\label{sec:concl}
This study extends the surface flux transport (SFT) optimisation of \citet{petrovay2019optimization} by incorporating explicit nonlinear quenching mechanisms, TQ and LQ, together with a tunable flux-decay term. The inclusion of this feedback significantly refines the admissible parameter space of $(u_{0}, \eta, \tau)$, transforming the broad islands of solutions identified in earlier studies into narrow, physically consistent ridges. The results demonstrate that LQ exerts a stronger influence than TQ, producing a pronounced ceiling effect that limits the amplification of the axial dipole moment in strong cycles. The optimal decay timescale is found to lie near $\tau \simeq 8$--10~yr, consistent with the physically realistic range inferred from previous optimisation studies.

The combined action of TQ, LQ, and flux decay establishes a tightly coupled balance between advection, diffusion, and flux loss. This coupling naturally leads to a self-limiting, weakly nonlinear regime in which the solar cycle amplitude is saturated by feedback from the surface field itself. The admissible domains revealed by the optimisation suggest that the Sun operates close to this saturation threshold, where small perturbations in surface transport or emergence properties can lead to substantial modulation in polar field strength. Consequently, these nonlinear feedbacks inherently limit the predictability of the solar cycle: while the mean amplitude and timing remain constrained by the quenching-regulated balance, stochastic fluctuations in active region emergence or flow perturbations can still drive cycle-to-cycle variability.

The results further suggest a physical connection between latitude-dependent quenching and the inflows toward active regions observed in Doppler and helioseismic measurements. As demonstrated by \citet{talafha2025effect}, these inflows act as an additional nonlinear regulator that suppresses flux emergence at low latitudes, effectively reinforcing the LQ mechanism included in this work. Such surface inflows, together with global quenching effects, provide a unified picture of how magnetic activity modulates its own transport environment, linking the SFT description at the surface to the deeper dynamo processes operating below.

In summary, the incorporation of nonlinear quenching yields a more constrained and physically self-consistent optimisation of SFT parameters. The emerging picture is that of a dynamo system operating in a marginally stable regime, where nonlinear feedbacks—acting through both surface and subsurface processes—maintain solar cycle amplitudes within a narrow range while allowing for moderate, stochastic variability. Future extensions of this framework will include data assimilation of WSO magnetograms and the explicit treatment of cycle-dependent inflows, enabling direct observational calibration of the quenching parameters and providing improved constraints for predictive dynamo modelling.

\begin{acknowledgements}
This research was funded by the College of Graduate Studies and supported by the Research Institute of Sciences and Engineering (RISE) at the University of Sharjah. This research acknowledges the use of synoptic magnetic field data provided by the Wilcox Solar Observatory (WSO). 
\end{acknowledgements}

\bibliographystyle{aa} 
\bibliography{reference}
\end{document}